# Self-Stimulated Undulator Radiation and its Possible Applications

E.G.Bessonov[†], M.V.Gorbunkov[†], A.A.Mikhailichenko[††], A.L.Osipov[†], A.V.Vinogradov[†]

[†]Lebedev Phys. Inst. RAS, Moscow, Russia
[††]Cornell University, LEPP, Ithaca, NY 14853, U.S.A.

*Abstract:* The phenomena of self-stimulation of incoherent radiation emitted by particles in a system of undulators installed in the linear accelerators or quasi-isochronous storage rings is investigated. Possible applications of these phenomena for the beam physics and light sources are discussed.

## 1. Introduction

Self-Stimulated Undulator Radiation (SSUR) is a kind of radiation emitted by a charged particle in a field of the downstream undulator in the presence of self-fields of its own wavelets emitted at earlier times in the same or upstream undulator. These wavelets focused back to the particle's position at the entrance of the downstream undulator with mirrors, lenses and passed through the optical delay lines [1]. Numerous schemes of SSUR production can be suggested in general. Below we considering two of such schemes based on the magnetic lattices of the linear accelerators and storage rings. Requirements to the parameters of electron or ion beams (energy spread, emittance), magnetic lattices and degree of synchronicity are evaluated. The storage rings (including the compact ones) using the ordinary or laser undulators for generation of continuous, quasi-monochromatic radiation in the optical to X-ray regions are considered for usage of SSUR.

## 2. SSUR source based on the linear system of undulators

A particle passing through an undulator emits an undulator radiation wavelet (URW), the length of which in the direction of its average velocity is $M\lambda_1$ where $M$ is the number of undulator periods, and $\lambda_1$ is the wavelength of the first harmonic. In a system of $N_u$ identical undulators, located along straight line, the particle radiates $N_u$ URWs with a separation $l$; both $l$ and $\lambda_1$ are defined by the Doppler effect, by an angle $\theta$ between the average particle velocity in the undulator and the direction to the observer, by the distance between the undulators $l_0$, by the period of the undulator $\lambda_u$ and by the relativistic factor $\gamma \gg 1$. In the forward direction ($\theta = 0$) these numbers are: $l = l_0/2\gamma^2$ and $\lambda_1 \cong \lambda_u/2\gamma^2$. The energy radiated by a particle in a system of $N_u$ undulators becomes modified and is $N_u$ times larger than the one radiated in just a single undulator. The spectrum of radiation emitted in an arbitrary direction also changes, appearing as a line structure. The integrated spectrum does not change much however.

In the publication [1] a way to increase the loss rate of a particle in a system of $N_u$ undulators by the introduction of controlled delays in the motion of the particles between undulators relative to their URWs was suggested; see Fig.1.

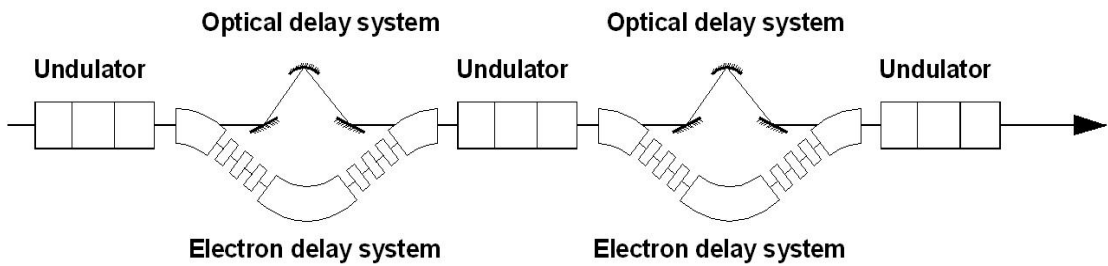

Figure 1: Schematic of the installation.

Delays are chosen so that a particle enters the following undulator in the decelerating phase at the front edge of its URW, which was emitted from preceding one. In this case the particle will experience deceleration in its self-field generated by its instantaneous motion in the field of the undulator (friction force generated by the spontantenous incoherent radiation) as well as in the field of the URW from preceding undulators (stimulated radiation in field of a co-propagating URW). Under such conditions superposition of the wavelets occurs, which yields the electric field growth $\sim N_u$ and the growth of energy density in the emitted radiation becomes $\sim N_u^2$. Below we will name the linear system of undulators by self stimulated undulator klystron (SSUK).

To be optimally effective, this system must use appropriate focusing elements such as lenses and/or focusing mirrors, see the schematics of installation in Figs. 1, 2. Mirrors and lenses are used to form a crossover in the middle of the undulators with the Rayleigh length of the order of the length of undulator $Z_R \cong M\lambda_u / 2$.

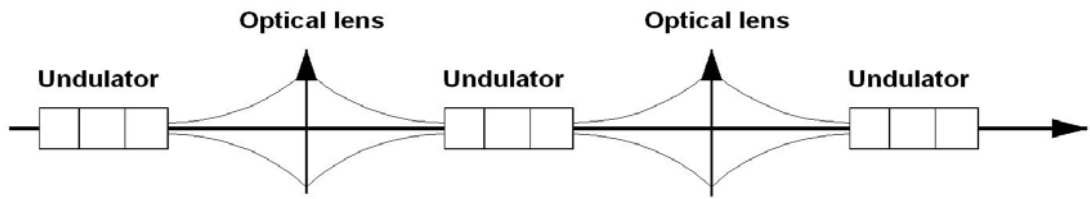

Figure 2: Equivalent optical scheme.

A reminder here is that for any method of Optical Stochastic Cooling (OSC) it is important to inject as many photons as possible in the spectral bandwidth $\Delta\lambda / \lambda \sim 1/2M$ and in the angles $\Delta\theta \sim 1/\gamma\sqrt{M}$ [2], [3], [4], [5]. The number of photons within this angular and energy spread does not depend on the length of the undulator. So, for example, the use of three pickup undulators in SSUK is 3 times more effective in the emitted field strengths and 9 times more effective in the emitted energy (in the number of emitted photons) than just in a single pickup undulator. This also means that usage of such system with three pickup undulators and a single kicker for OSC is 3 times more effective for damping, than just a single pickup and a single kicker. So the effectiveness of the pickup and kicker SSUK system consisting of $N_u$ undulators each is proportional to $N_u^2$.

We considered here the case where the optical delays are tuned so that the wavelets emitted by the particle are congruent and the particle always stays at the decelerating phase. For this the beam delay system must be isochronous for all particles in the beam selected for cooling.

Another important item is the destructive interference with URWs radiated by other particles. This process is similar to OSC, leading to reduction of cooling process proportional to $1/\sqrt{N_{bw}}$, where $N_{bw} \cong NM\lambda_1 / l_b$, stands for the number of particles in the bandwidth, $l_b$ is the bunch length and $N$ is its population.

We have considered a case, in which the URWs are emitted by each particle individually. The same consideration still valid for the single micro bunch with the number of particles $N_1$ and the length $l_{mkb} << \lambda_1$ (such micro bunch is equivalent to a particle with the charge $eN_1$) or for the trains $N_{mkb}$ such micro bunches in a parametric (prebunched) Free Electron Lasers (FEL) [6]. In this case the power emitted is $P \sim N_1^2 N_u^2$. The system which consist of a modulator undulator with the laser beam and the system of radiator undulators installed in a storage ring (as well as in ordinary or energy recovering linacs and recirculators) can be used by analogy with the scheme considered in [7].

## 3. SSUR source based on storage rings

The SSUR source is based on a quasi-isochronous storage ring equipped with an undulator installed in its straight section and the mirrors installed at both sides of undulators outside of the closed orbits of electrons, circulating in the ring (Fig.3). So the mirrors set an optical resonator.

The scheme of the SSUR source has resemblance to the scheme of ordinary FEL with additional synchronicity condition: the oscillation period of the URW emitted by every electron in the undulator inside the optical cavity coincides with the revolution period of this electron in the storage ring for the large range of energy and transverse emittance of the beam. The URWs emitted by every electron are accumulated effectively in the optical resonator by the superposition one by another if theirs longitudinal shift per turn satisfies condition

$$|\Delta l| \leq \lambda_m / F, \qquad (1)$$

where $\lambda_m = \lambda_1 / m$ is the wavelength of the UR emitted by the electron on the $m$-th harmonic in the direction of its average velocity, $F$ is the finesse (quality factor) of the optical resonator.

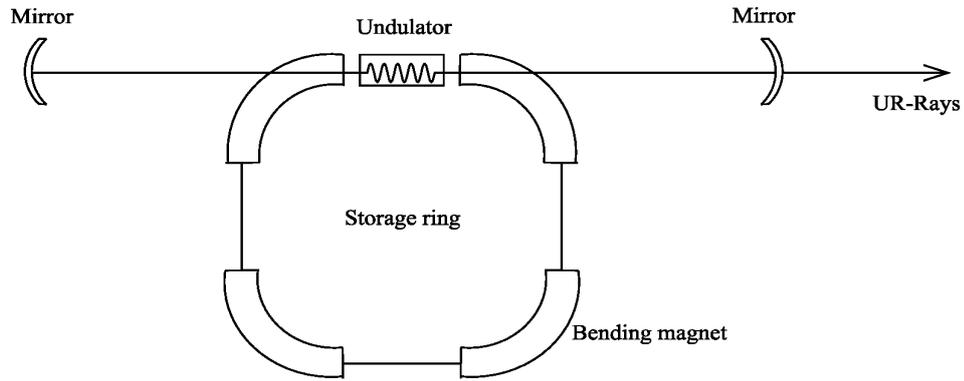

Fig.3. Schematic diagram of SSUR source built around a storage ring.

One important peculiarity of the source suggested here is that there is no requirement for the coherence in radiation among different electrons in the bunch like it is required for the pre-bunched FELs [8], [9], [10] including ones based on isochronous storage rings, [11], [12]. Electrons in this source are not grouped in micro-bunches with the longitudinal dimension $\sigma_\parallel \ll \lambda_m$, separated by the distances which are integers of $\lambda_m$. Stimulated process of radiation for each electron is going in the undulator with their own URW fields only. Every electron enters the undulator together with its URWs emitted at the earlier times [1].

The condition (1) presents the main synchronicity condition. In general case there are $2M+1$ similar collateral synchronicity conditions corresponding to incomplete overlapping of the URWs

$$|\Delta l - n\lambda_m| \leq \lambda_m / F, \qquad (2)$$

where $n = 0, \pm 1, \pm 2, ..., |n| \leq M$, $M$ is the number of the undulator periods.

Obviously, all properties of the spontaneous incoherent radiation emitted by the electrons in a SSUR source under main synchronicity condition $n = 0$ are not changed, except intensity, which becomes higher by $F/2\pi$ times. At collateral synchronicity conditions ($|n| > 0$) the URWs emitted by an electron at each pass through the undulator are shifted by the distances $\pm \lambda_m, \pm 2\lambda_m$, ... $\pm n\lambda_m$, ... $\pm M\lambda_m/2$ with the gaps $\Delta l \leq \lambda_m / F$ for the next URW relative to previous one and the properties of radiation are different: the intensity is dropped, but the monochromaticity is increased with the number $n$.

Note that the emitted wavelength of URW $\lambda_m \simeq \lambda_u(1+K^2)/2m\gamma^2$ depends on the relative energy $\gamma = \varepsilon / mc^2$ and hence, on the number $n$ of synchronicity condition, where $K = \sqrt{|\vec{p}_\perp|^2} =$

$e\sqrt{|\vec{B}_\perp|^2}\lambda_u/2\pi m_e c^2$ is the deflection parameter of undulator, $\vec{p}_\perp = \gamma\vec{\beta}_\perp$, $\vec{\beta}_\perp = \vec{v}_\perp/c$ is the transverse relative electron velocity exited by the field of undulator, $\vec{B}_\perp$ is the transverse component of the magnetic field strength vector, $m_e$ is the rest electron mass, $\varepsilon$ is the electron energy [13].

## 4. Requirements to the electron beam parameters of the SSUR sources, based on the storage rings

The general synchronicity condition in the SSUR scheme is as the following

$$\Delta l = |c \cdot \Delta T_{e,URW}| \ll \lambda_m / F, \qquad (3)$$

where $\Delta T_{e,URW} = T_e - T_{URW}$ is the difference between the revolution periods of the electron in the storage ring and the UWR in the optical resonator. We are considering $T_{URW} = 2L_{mir}/c = const$, $T_e = T_e(\varepsilon, A_b)$, where $L_{mir}$ is the distance between mirrors, $A_b$ is the amplitude of the electron betatron oscillations. The value $\Delta T_{e,URW}$ can be presented in the form $\Delta T_{e,URW} = \Delta T_\eta + \Delta T_{A_b}$, where in the smooth approximation $\beta_{x,z} \simeq \overline{\beta}_{x,z} = C/2\pi\nu_{x,z}$

$$\Delta T_\eta = \eta_c \cdot T \cdot \Delta\varepsilon/\varepsilon, \qquad c \cdot \Delta T_A = \pi^2 A_{b,x,z}^2 \nu_x^2 / C, \qquad (4)$$

$\eta_c = 1/\gamma^2 - \alpha_c$ is the phase slip factor of the ring [14], [15], $C$ – is the circumference of the electron orbit, $A_{b,x,z}$, $\beta_{x,z}$, $\nu_{x,z}$ are the horizontal/vertical amplitudes, $\beta$ -are the functions and tunes of electron betatron oscillations accordingly, $\alpha_c$ is the momentum compaction factor of the ring [15]. For relativistic electron beams $\eta_c \simeq -\alpha_c$. Synchronicity condition determines the limiting energy spread, amplitudes of betatron oscillations and emittance of the electron beam:

$$\Delta\varepsilon_r/\varepsilon < \lambda_m/CF\eta_c, \qquad A_{b,x,z} < \sqrt{\lambda_m \lambda_{x,z}/F\nu_{x,z}}/\pi, \qquad \in_{x,z} < 2\lambda_m/\pi F\nu_{x,z}, \qquad (5)$$

where $\lambda_{x,z} = C/\nu_{x,z}$ is the wavelength of the betatron oscillations. Note that the last inequality in (5) $\nu_{x,z} F/4 > 1$ times stronger than one for the diffraction limited electron beam. The used smooth approximation permits to appreciate the expressions for requirements to the storage ring.

In order to URWs emitted in the direction of the undulator axis overlapped effectively they must have small spread of the carrier frequency. It follows from here that the requirements to the energy and angular spreads of the electron beam and its emittance should be

$$\Delta\varepsilon/\varepsilon \ll 1/m \cdot M, \qquad \Delta\theta \ll 1/\gamma\sqrt{m \cdot M}, \qquad \varepsilon_{x,z} = \overline{\beta}_{x,z} \cdot \Delta\theta^2 < \lambda_m \lambda_{x,z}/\pi L_u, \qquad (6)$$

where $L_u = M\lambda_u$ is the undulator length. Usually the values $\lambda_m/C \cdot F \cdot \eta_c \ll 1/mM$, $F \gg 2L_u/C$. That is why the conditions (6) are less severe than the (5).

Note, that if the energy spread of the beam $\Delta\varepsilon_b$ is much bigger than the limiting one $\Delta\varepsilon_r$, then $\sim 2M+1$ collateral synchronicity conditions $\Delta l_\eta = c\Delta T_\eta = \pm n\lambda_m$ can occur simultaneously at the energies determined by the different numbers $n$ with the limiting energy spreads determined by (5). For $|n|=M$ the acquired relative energy spread of the beam $2MF$ times bigger:

$$\Delta\varepsilon_b/\varepsilon < 2M\lambda_m/C\eta_c. \qquad (7)$$

The requirements to the beam emittance for the collateral synchronicity conditions stay (5). They are less severe than requirement for the limiting energy spread $\Delta\varepsilon_r$.

*Example.* Let $\lambda_m = \lambda_1 = 1$ mkm, $C$=100 m, $\eta_c = 10^{-5}$, $\nu_x = 2.5$, $F = 31.4$, $M$=30. In this example, according to (5), the electron beam for the main synchronicity condition must have the emittance $\in_{x,z} < 8.1$ *nm,* energy spread $\Delta\varepsilon_r/\varepsilon \ll 3.2 \cdot 10^{-5}$, $\Delta\varepsilon_b/\varepsilon \ll 3.01 \cdot 10^{-2}$ and amplitudes $A_{z,x} \ll 0.22$ mm.

The requirements to the electron beam emittance in the optical region are acceptable for the 2[nd] Generation Light Sources (GLSs). At the same time the requirements to the electron beam energy spread $\Delta\varepsilon_r/\varepsilon$ are severe for the modern 3$^{rd}$, future 3.5$^{th}$ Generation Light Sources (GLSs) [16] and for the LSs using multi-turn recirculation of the ultra low emittance electron beams in the energy recovery storage rings [17], [18], [19], [20]. The energy spread and the emittance of the beam can be decreased if the energy and the current of the beam in the storage rings could be decreased.

If we accept, that the normalized transverse beam emittance is $\epsilon_n = 0.1$ mm·mrad and the relative energy spread is $\Delta\varepsilon_b/\varepsilon \ll 10^{-4}$ then at the energy $\varepsilon = 5$ GeV ($\gamma = 10^4$) the geometrical emittance will be equal to $\epsilon = 10^{-11}$ m. These beam parameters are typical for the advanced 4$^{th}$ GLSs under development and are one order of magnitude bigger than for 3$^{rd}$ GLSs [21].

Requirements to the electron beam energy spread and emittance are increased with the hardness of the UR $\hbar\omega_m = 2\pi\hbar c/\lambda_m$. It is difficult to obtain the energy spread of the electron beam $\Delta\varepsilon_r/\varepsilon$ necessary for synchronicity conditions in the X-ray region ($\lambda_m \sim 1 A^0$), but it is possible to obtain the energy spread $\Delta\varepsilon_b/\varepsilon$ and to work with the main and collateral synchronicity conditions ($0 < |n| < M$). Some electrons in the beam will work effectively in this case, if the requirements (1), (2), (5) are satisfied.

### 5. To the beam dynamics in a quasi-isochronous storage ring

One of the main problems in the considered scheme of the SSUR source is the condition of synchronicity(3), which can be realized in an isochronous storage ring ($|\eta_c| \ll 1$). In a year 2007 Helmut Wiedemann wrote: "An electron storage ring where the momentum compaction is adjusted to be zero or close to zero is called an isochronous ring or a quasi-isochronous ring. Such rings do not yet exist at this time but are intensely studied and problems are being solved in view of great benefits for research in high energy physics, synchrotron radiation sources, and free electron lasers to produce short electron or light pulses." (see [15], p. 303). To the present day all projects and experiments were done with existing storage rings converted to isochronous ones by some changes in theirs lattices. The problem of optimal isochronous storage ring for any given purpose is not solved. This is a difficult problem. D. S. Robin and W. Wan for such case in 2007 wrote: "Let's assume that one has a storage ring and wants to adjust the lattice settings to obtain certain properties such as low emittance or a small momentum compaction, etc. Determining how to adjust the lattice to achieve certain properties is not a straight forward process. The process is actually blind and involves a lot of trial and error. In many ways it is an art which is aided by the instincts and experience of the practitioner" [22].

To understand the problem of the isochronous storage ring, we should investigate the auto phasing problem of the electrons when slip factor $\eta_c$ is small. The principle of phase focusing is a fundamental one for the beam stability in the circular accelerators and storage rings. It is described by the system of equations

$$\frac{d\varphi}{dt} = -c\beta q \eta_c \frac{\Delta p}{p_s}, \qquad \beta\Delta p = \frac{\Delta\varepsilon}{c}, \qquad \frac{d\varepsilon}{dt} = \frac{\Delta\varepsilon_{turn}}{T_s} = eV_0 \sin\phi - U_s \qquad (8)$$

where $\varphi$ is the electron phase, $\beta = v/c$ is the relative electron velocity, $q$ is harmonic order, $p = m_e c\beta\gamma$ is the electron momentum, $T_s$, $U_s$ are the revolution period and the frictional energy losses of synchronous electron per turn, $V_0$ is the amplitude of the RF voltage [23]. Damping and fluctuation terms in equation (8) are neglected.

The slip factor determines critically the beam stability. If the slip factor is reduced to zero, then, according to (8), the phase of electron is not changed, its energy is going in the same direction up or down depending on the position of the electron phase $\varphi$ relative to the synchronous phase $\varphi_s$. The linear theory of phase focusing predicts the beam loss in such case due to lack of phase stability. In this case or if the slip factor is very small, the higher order terms in the expansion of the slip factor over the energy and amplitudes of betatron oscillations become significant.

The problems associated with isochronous storage rings can be facilitated if the phase focusing rejected and the ring forced to work with low-friction particles (ion and muon beams, low energy electron beam) and the eddy fields or phase displacement mechanisms. In this case the beam will be unbunched. Here we would like to underline that the lattice can be constructed in a such way that the slip factor will reach zero value, cross it or will have minimum at given energy of the electron. Near to these amplitudes and energies the regions of the energy and amplitudes (5) appear which satisfy the synchronicity condition (3). The problem is in development of a lattice for the quasi-isochronous storage ring, which has zero slip factor at given transition energy and the revolution time of electrons near the transition energy has weak dependence on their momentum and amplitudes of betatron oscillations.

## 6. Excitation of an optical resonator by the electron beam

The electric field strength of the circularly polarized Gaussian laser beam in an open resonator for the fundamental transverse TEM$_{00}$ mode is given by

$$\vec{E}_\lambda(r,z,t) = E_{0,\lambda} \frac{w_0}{w(z)} \exp(\frac{-r^2}{w^2(z)})[\vec{e}_x \cos(\omega_\lambda t - k_\lambda z - \frac{kr^2}{2R} + \varsigma(z)) +$$

$$\vec{e}_y \sin(\omega_\lambda t - k_\lambda z - \frac{kr^2}{2R} + \varsigma(z))], \qquad (9)$$

where $r$ is the radial distance from the central axis of the beam, $z$ is the axial distance from the beam's waist, $k = \omega/c = 2\pi/\lambda$ is the wave number, $\lambda$ is the wavelength, $E_{0,\lambda} = E_\lambda|_{r=z=0}$, $w(z) = w_0\sqrt{1 + z^2/Z_R^2}$ is the radius at which the field amplitude and intensity drops to $1/e$ and $1/e^2$ of their axial values, respectively, $w_0 = w(0)$ is the waist size, $Z_R = \pi w_0^2/\lambda$ is the Rayleigh length, $R(z) = z[1 + Z_R^2/z^2]$, $\varsigma(z) = arctg(z/Z_R)$. The corresponding time-averaged intensity is

$$I_\lambda(r,z) = I_{0,\lambda}\left(\frac{w_0}{w(z)}\right)^2 \exp\left(-\frac{2r^2}{w^2(z)}\right), \qquad (10)$$

where $I_{0,\lambda} = I_\lambda(0,0)$ is the intensity at the center of the beam at its waist. The same form (9) has the vector potential $\vec{A}_\lambda(r,z,t)$. Its components can be presented in the form $\vec{A}_{\lambda,x,y}(\vec{r},t) = \vec{e}_{x,y} q_{\lambda,x,y}(t)A_\lambda(\vec{r})$, where $q_\lambda(t)$ is the amplitude of the vector potential and $\vec{A}_\lambda(\vec{r})$ is the eigenfunction of the resonator normalized by the condition $\int_V |\vec{A}_\lambda(\vec{r})|^2 dV = 1$. The total free electromagnetic field in the resonator is described by the expression $\vec{A}(\vec{r},t) = \sum_\lambda [\vec{e}_x q_{\lambda,x}(t) + \vec{e}_y q_{\lambda,y}(t)]A_\lambda(\vec{r})$. The electric field strength in the EM wave which is not restricted transversely in the resonator, can be expressed through the vector potential in the form $\vec{E}_\lambda(\vec{r},t) = \dot{\vec{A}}_\lambda(\vec{r},t)/c = k_\lambda q_\lambda \vec{A}_\lambda(\vec{r},t)$. The normalization condition in this case leads to $\int_V |\vec{E}_\lambda(\vec{r})|^2 dV = k_\lambda^2 = \pi w_0^2 L_{mir} E_0^2/2$. It follows from here, that the normalized amplitudes are $A_0 = 1/w_0\sqrt{\pi L_{mir}/2}$, $E_0 = k_\lambda/w_0\sqrt{\pi L/2}$.

The equation for the amplitude of the eigenmode $q_\lambda$ for free fields in the resonator is the following [24]

$$\ddot{q}_\lambda + \frac{\omega_r}{F_\lambda}\dot{q}_\lambda + \omega_\lambda^2 q_\lambda = \int_V \vec{J}(\vec{r},t)\vec{A}_\lambda(\vec{r},t)dV \tag{11}$$

where $F_\lambda$ is the resonator finesse related to the frequency $\omega_\lambda$, and the dot means time derivative. This is the equation of damped harmonic oscillator excited by the external force $f(t) = \int_V \vec{J}(\vec{r},t)\vec{A}_\lambda(\vec{r},t)dV$. The energy in every mode appears instantaneously with the electron entering the undulator and is growing in time in accordance with the electron motion and radiation in the resonator. After the electron exits the resonator the energy in the modes stay constant. In reality, we must summarize the electromagnetic fields in the resonator. Only in this case we will obtain the compact URW in the resonator with no energy located outside of the URW.

Below we simplify the problem and accept that the URW is in a steady-state regime under the synchronicity condition $\Delta l = 0$. The harmonic $m=1$ is excited in a helical undulator. In this case the circularly polarized electric field strength of the emitted URW has the form (9) with the amplitude $E_0 = E|_{r=z=0}$, the stored energy of the URW is $\varepsilon_{URW} = \int_V (E^2/4\pi)dV = E_0^2 w_0^2 M\lambda_1/8$, where $M\lambda_1$ is the length of the URW. For the steady-state regime the energy loss of the URW per single turn is equal to the energy loss of the electron in the fields of undulator and the URW. For single electron $\vec{J}(\vec{r},t) = e\vec{v}_e(t)\delta[\vec{r}-\vec{r}_e(t)]$, where $\vec{r}_e(t)$, $\vec{v}_e(t)$ are the electron radius-vector and velocity respectively. In this case the energy emitted by the electron in the URW per single pass is $\Delta\varepsilon_e = \int_0^T \int_V \vec{J}(\vec{r},t)\vec{E}_{URW}(\vec{r},t)dVdt = e\int_0^T \vec{v}_e(\vec{r}_e,t)\vec{E}_{URW}(\vec{r}_e,t)dt = e\beta_{e,\perp}E_0 M\lambda_u$. The damping of the URW energy for single revolution is $\Delta\varepsilon_{URW} = \varepsilon_{URW}[1-\exp(-\delta T_{URW})] \simeq \pi w_0^2 M\lambda_1 E_0^2/4F$, where $\delta$ is the damping decrement of the URW, $1-\exp(-\delta T_{URW}) \simeq \delta T_{URW} \simeq 2\pi/F$. From the energy balance $\Delta\varepsilon_e = \Delta\varepsilon_{URW}$, the accepted condition $Z_R = M\lambda_u/2$ or $w_0^2 = M\lambda_u\lambda_1/2\pi$ and in case of high finesse $F/2\pi \gg 1$ the electric field strength $E_0 = 8e\beta_\perp F/M\lambda_1^2$, the energy of the URW $\varepsilon_{URW} = 8e^2 K_\perp^2 F^2/\pi\lambda_1(1+K^2)$. It follows from here that the amplitude of the URW emitted by the electron per pass is $\Delta E_0 = 2\pi E_0/F = 16\pi e\beta_\perp/M\lambda_1^2$, the energy of the URW and the number of the photons emitted for a single pass are

$$\Delta\varepsilon_{URW,1} = 32\pi e^2 K_\perp^2/\lambda_1(1+K^2), \quad N_\gamma = 16\alpha K^2/(1+K^2) \tag{12}$$

where $\alpha = e^2/\hbar c = 1/137$. Non-synchronous condition of excitation of the resonator $\Delta l \neq 0$ can be investigated by analogy with excitation of resonators by the periodic electron bunches in the parametric FELs [25]. Note that in this case behavior of the energy variation of the URW in the optical resonator is similar to the energy dependence of an oscillator excited by an external force (see Appendix).

Note that in many applications the beam intensity is determined by the rms beam size $\sigma(z)$ with the form

$$I_\lambda(r,z) = I_{0,\lambda}\left(\frac{\sigma_0}{\sigma(z)}\right)^2 \exp\left(-\frac{r^2}{2\sigma^2(z)}\right). \tag{13}$$

According to (10) and (13) $w(z) = 4\sigma(z)$ and $Z_R = 4\pi\sigma^2(z)/\lambda$.

### 7. Frictional cooling of electron and ion beams

The beam cooling associated with increase of six-dimensional (6D) phase space density of the beam due to reduction of its emittance. There are two ways to the particle beam cooling. They are based either on friction (frictional force is directed against the particle velocity) or on the limiting the inter-particle spacing (follow the particle of the beam and force the particle to shift to the center of the beam) [26], [27]. In any case personal force must be applied to each particle individually. External fields and self-fields produced by particle beam do not lead to cooling as they act upon all particles around the reference one.

According to the generalized Robinson damping criterion for the frictional cooling, the rate of the 6D beam density change is determined by the damping decrement

$$\alpha_{6D} = (1+\frac{1}{\beta^2})\frac{\overline{P}_{Fr}(\varepsilon)}{\varepsilon} + \frac{\partial \overline{P}_{Fr}(\varepsilon)}{\partial \varepsilon}, \qquad (14)$$

where $\overline{P}_{Fr}(\varepsilon)$ is the power of the electromagnetic energy emitted by the particle [27]. Different parts of the beam in the phase space volume occupied by the beam can have different rates of cooling. In case of SSUR, the partial derivative of the power $\partial \overline{P}_{Fr}(\varepsilon)/\partial \varepsilon$ strongly depends on the deviation of the particle's energy on the energy corresponding to the condition of main or collateral synchronicity.

If the resonator is switched off, then, according to (12), about 0.1 equivalent photons per pass with the energy $\sim \hbar \omega_1$ is emitted in the undulator with the undulator deflection parameter $K>1$. In this case the photon energy and the frictional power $\overline{P}_{Fr}(\varepsilon) \sim \varepsilon^2$. That is why in the relativistic case the first and second terms in (14) are equal. If the resonator is turned on, the power $P_{Fr}(\varepsilon,t)$ at the energy corresponding to the synchronicity condition is increased $\sim F/2\pi$ times and, according to (5), (6), the partial derivative $\partial \overline{P}_{Fr}(\varepsilon)/\partial \varepsilon$ at the bias of the dependence $\overline{P}_{Fr}(\varepsilon)$ is increased $\sim (F/2\pi)(\lambda_m M / CF\eta_c) = \lambda_m M / 2\pi C\eta_c$ times. That is why the second term in (14) will be $\sim \lambda_m M / 2\pi C\eta_c$ times higher than the first one and the damping time will be $\sim F\lambda_m M / 2\pi C\eta_c$ times smaller.

If we introduce a delay line in the resonator and amplifier, then the possibility appears to move the energy corresponding to the synchronicity condition down and to gather the dense particle beams at the lower energy (the analogy of the frequency chirp for the laser cooling of ion beams). The efficiency of this scheme can be increased if we will turn on the amplifier per duration of the particle bunch for damping time of the URWs, then turn it off, removing by this way the stored laser energy from the resonator per one revolution and repeat this process many times. In this case the damping process can be many orders of magnitude shorter. Different schemes of cooling can be considered (RF accelerating fields are switched on/off, eddy fields or phase displacement mechanisms are used, their combinations). IR, optical and UV amplifiers are available for SSUR cooling of particle beams.

Notice, that the frictional cooling supposes individual action of the friction force on a particle. A particle moving non-uniformly in the external fields produces the self fields (both Coulomb, longitudinally and transversely) distributed in space. These fields have both transverse and longitudinal components relative to the particle velocity. They form the frictional forces for the considered particle and stochastic disturbing forces for the neighboring electrons. Joint fields of the particle beam do not violate Robinson damping criterion. But individual interaction of particles can lead to excitation of betatron and phase oscillations not only through the Coulomb fields (intrabeam scattering) but through the transverse ones as well.

## 8. Light Sources based on self-stimulated undulator radiation

The scheme of the SSUR source requires an electron beam with ultralow transverse emittance, energy spread and an optical resonator with high finesse (quality factor). This is possible in cm to optical and UV regions. Very high finesse (above $10^6$) can be achieved either by using the dielectric super-mirrors or in certain microcavities based on whispering gallery modes [28]. The problem of X-ray mirrors is not solved and stays very important one. The only versions for such mirrors applicable for the Light Sources (LS) now are the mirrors based on the Bragg scattering [29], [30], [31], [32], [33], [34]. These mirrors effectively reflect radiation in a very narrow spec-tral range. For normal incidence the reflection of X-rays from the diamond under the Bragg condition could approach 100% - substantially higher than for any other crystal. Commercially produced synthetic diamond crystals demonstrate an unprecedented reflecting power at normal incidence and millielectronvolt-narrow reflection bandwidths for hard X-rays [35]. Electron beams with normalized emittance *1 mm mrad* exist now. One order smaller emittances are under discussion.

SSUR source in the limiting case $\Delta\varepsilon_b / \varepsilon = 0$, $\eta_c = 0$, $\in = 0$ and under the general synchronicity condition ($n=0$, total overlapping of URWs for all electrons of the beam) will have the power which is $F/2\pi$ times higher than spontaneous incoherent radiation acquired in the resonator with TEM$_{00}$ mode outside of the synchronicity condition.

In this case under the collateral synchronicity condition corresponding to the contact between the neighboring URWs *(|n|=M)*, the power of the SSUR will be equal to the incoherent one but the monochromaticity will be increased $F/2\pi$ times (similar to the case of parametric FELs [25]), the length of the effective URW will be equal $l_{URW}^{ef} \simeq F l_{URW} / 2\pi \simeq M F \lambda_1 / 2\pi$. If in this case the length of the electron bunch $l_b \ll l_{URW}^{ef}$, then both every emitted URW and the total bunch of the UR (the sum of the emitted URWs) will be described (except a part of the length $l_b$ for total bunch) by approximately pure sine wave with slowly decreased amplitude. The power in the total URW is equal to the power of the spontaneous incoherent radiation. Note that the spontaneous incoherent UR consists of the large number of short independent URWs and usually $l_b \gg l_{URW}$. In this case there is no phase correlation between the URWs and the form of the total electric field strength in the UR bunch is far from sine like one.

If the smallness conditions (5) for the energy spread, the slip factor and the emittance are violated, then in the 6D phase space region occupied by the electron beam ~$2M$ sub-regions satisfied to the synchronicity conditions (2) appear. However the effect on the total power amplification for such beam will not be high because of the phase space volume occupied by sub-regions is less then the total 6D volume. The transient behavior of the power of the emitted undulator radiation can be used for its amplification (see Appendix).

By using a linear system of $N_u$ pickup undulators, SSUK, (see Figures 1, 2) located along the straight section of the storage ring one can amplify this process $N_u^2$ times. Note that for the SSUK the requirements to the bunch parameters are much easier then (5). They are determined by (5) if we replace C on $l_0$, $\eta_c$ on $\eta_{c,l}$, $\lambda_{x,z}$ on $l_0$, and suppose $\nu_{x,z} = F = 1$, where $\eta_{c,l}$ is the local slip factor do not burden by the auto phasing problem [2]. There is no problems to produce the isochronous linear system of undulators with the zero local slip factor [36].

The Bragg reflecting crystal mirrors have a peculiarity. They absorb a small fraction of energy in the small frequency range $\Delta\omega_{refl}$ ($\hbar\Delta\omega_{refl} \sim 1\ meV$) during URWs round trip. The intensity reflection coefficient in this frequency range can be high $r_{Br} \simeq 1 - 2\pi / F_{refl} \simeq 0.99$ and near to zero ($r_{Br} \simeq 0$) in the rest spectral region. The total energy of the URWs during round trip reflection will be decreased $r_{Br} r_{Br,tot}^{-1} = \Delta t_{refl} / \Delta t_{URW} = M_{refl} / M \gg 1$ times, where $\Delta t_{refl} = T_1 M_{refl}$ is the duration of the reflected URW, $T_1 = 2\pi / \omega_1$, $M_{refl} = \hbar\omega_1 / \Delta(\hbar\omega)_{refl} \simeq 10^6 - 10^7$ is the number of cycles in the reflected URW. The degree of monochromaticity $\Delta\omega / \omega$ and coherence length of the reflected URW $l_{refl} = c\Delta t_{refl}$ will be increased $r_{Br,tot}^{-1}$ times.

The fronts of URWs reflected by Bragg mirrors will coincide with the initial ones. Electrons will emit their URWs at different moments of time in the limits of the electron bunch current duration $\Delta t_b$. The lengths of the reflected URWs $l_{refl} = \lambda_1 M_{refl}$ can be much larger than the bunch length $l_b = c\Delta t_b$. In this case the UR bunch after reflections in the resonator will be presented by one long ( $l_{refl} \gg l_b$ ) nearly pure sine wave except short (~$l_b$) head and tail parts of the beam.

X-ray version of SSUR source based on the Bragg reflecting crystal mirrors could be effective if quasi-isochronous storage rings, ultralow emittance electron or cooled ion beams and high finesse mirrors could be used. Backward Compton scattering sources based on compact lattices and laser undulators can be discussed as well.

## 9. SSUR sources and Free-Electron Lasers

When we are talking about a quasi-isochronous storage ring we have in mind that round trip slip factor of the ring is set to zero. At the same time the local slippage factor can be high in the region occupied by the undulators. It means that bunching of the beam and the emission of coherent UR can be produced by external electromagnetic wave in the undulator or in the SSUK (undulator/optical klystron mechanism [37], [38], [39]). If the large number of electrons satisfying to the synchronicity conditions (2) are located on the length of the URWs $M\lambda_1$ (coherence length, sample) then stimulation of SASE regime by high value seeding URWs from sub-regions satisfied to the synchronicity conditions will appear. Self-bunching will appear as well. In this case outside the undulator or SSUK the bunching can be lost but it will appear again and will be amplified in addition to the previous one by stored co-propagated URWs for the next turns through the same undulator or SSUK. By such way stimulation of the oscillator X-ray free-Electron laser regime under the main and collateral synchronicity conditions can be produced.

Unfortunately in the case of Bragg resonators the amplification of the reflected UR beam will be done for the very small part of the beam being amplified (for the electron beam duration $\Delta t_b$). At the exit of the undulator the seed URW will be amplified to the extent of $(1+g)r_{Br}r_{Br,tot}I_{inc}$ for the duration $\Delta t_b$, where $g$ is the gain of the FEL, $I_{inc}$ is the initial incoherent intensity of the UR beam emitted after first pass of the electron beam through undulator. Only a part $r_{refl,2} = \Delta t_b / T_1 M_{refl} \ll 1$ of the added radiation will be adopted by the Bragg resonator. That is why in this case the threshold condition is $(1+gr_{refl,2})r_{Br}r_{Br,tot} \gg 1$ [33]. This is very hard condition, however. Using SSUK can amplify this process $N_u^2$ times.

## 10. Conclusion

The phenomenon of self-stimulated incoherent emission of the UR in the SSUKs and quasi-isochronous storage rings is investigated. The requirements to the beam parameters (emittance, energy spread) and the degree of synchronicity are evaluated for the SSUR source based on a quasi-isochronous storage ring. SSUR source based on either the ordinary and compact storage rings using the static or laser undulators, electron or ion beams, ordinary or Bragg resonators are able to generate both short and continuous, quasi-monochromatic light beams in the optical to X-ray regions.

A transient behavior of the amplitude and the power of the URWs are investigated. It was shown that these values are the quasi-periodic functions of the revolution number in the time interval determined by damping time of the URW in optical resonator. At this interval the power of emitted radiation can be much higher than its steady state value. That is why the emitted power can be increased if the energy of optical beam stored in the resonator will be extracted periodically (or, if the phase of the stored radiation in the URWs will be changed to $\pi$) for one revolution of the beam in the ring (overload conditions).

The schematic SSUK could be used effectively in different methods of optical cooling (OC) of particle beams (ion, muon) in ordinary (non isochronous) damping rings [2], [3] - [5]. So these systems can serve as an effective pick-up undulator, for example. According to OC principle the optical parametric amplifier(s), controllable screens [2] and kicker undulators could be located in the subsequent straight sections.

SSUKs could be used effectively both in ordinary and prebunched FELs.

Excitation of resonators and a new scheme of cooling in the quasi-isochronous storage rings were discussed as well. Peculiarities of the SSUR emission can be used as an alternative method of muon and ion cooling in the storage rings.

This work was supported in part by RFBR under Grants No 09-02-00638a, 09-02-01190a.

## Appendix

The electric field strengths of an URWs emitted by a particle during its pass through the undulator on the *n-th* turn in the storage ring can be presented by the expression $E_n(t,n) = E_0 \sin[\omega_1(t-(n-1)\Delta t)]$ in the interval $[(n-1)\Delta t < t < n\Delta t]$, and $E_n(t,n) = 0$ outside of the interval, where $0 < t - (n-1)\Delta t < MT_1$, $\Delta t$ is the slip time, the moment $t = (n-1)T_{URW}$ corresponds to the arrival of the *n-th* URW to the observation point. The electric field strength of the sum of URWs emitted by a particle on its *n*-th pass of the undulator and URWs emitted by a particle on its previous 1,2,3… *(n-1)*-th passages of the undulator and reflected by mirrors of the optical resonator is

$$E(t,n) = E_0 \sum_{n=1}^{n} \sin[\omega_1(t-(n-1)\Delta t)] \, e^{-(n-1)\beta} , \tag{A1}$$

where we took into account the damping of the URWs by the coefficient $\beta = \delta T_{URW} = 2\pi / F$ determined by the mirror reflectivity $r = 1 - 2\pi / F$.

The expression (A1) is brought to the geometric progression

$$E(t,n) = E_0 \, \mathrm{Im}\{e^{i\omega_1 t} \sum_{n=1}^{n} e^{-(i\omega_1 \Delta t + \beta)(n-1)}\} = E_0 \, \mathrm{Im}\{e^{i\omega_1 t} \frac{1-e^{-(i\omega_1\Delta t+\beta)n}}{1-e^{-(i\omega_1\Delta t+\beta)}}\}.$$

which can be presented in the form $E(t,n) = E_0 a_n \sin(\omega_1 t) + E_0 b_n \cos(\omega_1 t)$ or

$$E(t,n) = E_n \sin(\omega_1 t + \xi_n) \tag{A2}$$

where $E_n = E_0 A_n$, $A_n = \sqrt{a_n^2 + b_n^2}$, $a_n = \dfrac{1 - e^{-\beta}\cos a - e^{-\beta n}\cos(an) + e^{-\beta(n+1)}\cos[a(n-1)]}{1 - 2e^{-\beta}\cos a + e^{-2\beta}}$,

$b_n = \dfrac{e^{-\beta}\sin\alpha - e^{-\beta n}\sin(\alpha n) + e^{-\beta(n+1)}\sin[\alpha(n-1)]}{1 - 2e^{-\beta}\cos a + e^{-2\beta}}$, $\xi_n = \arccos a_n / A_n$, $\alpha = \omega_1 \Delta t$,

$\beta = 2\pi / F$ is the damping decrement in units of the revolution number *n*.

As was expected if the value $\alpha = \Delta t = 0$ then $a_n = A_n = (1-e^{-\beta n})/(1-e^{-\beta})|_{\beta\to 0} \to n$, $a_n = A_n|_{\beta n \gg 1} = 1/(1-e^{-\beta})|_{\beta\to 0} = 1/\beta$, $b_n = \xi_n = 0$.

If $\alpha \neq 0$, the function $A_n$ is varied on the time interval $0 < t < MT_1 / \Delta t$ and come into steady state regime at $n > MT_1 / \Delta t$. In the steady state regime $A_n$ is determined by (A2) at the condition $n = n_c = MT_1 / \Delta t$.

It follows from (A2) that if $\beta \ll 1$ then the amplitude of the URW $E_n$ is the quasi-periodic function of the revolution number *n* with the angular frequency $\alpha$ (with the period $n_T = T_1 / \Delta t$) and with negative-going amplitude for the damping time $n_\tau = 1/\beta$. At $\beta n_c \gg 1$, $n > n_\tau$ it tend to the steady state regime:

$$a_n = \frac{1-e^{-\beta}\cos\alpha}{1-2e^{-\beta}\cos a + e^{-2\beta}}, \qquad b_n = \frac{e^{-\beta}\sin\alpha}{1-2e^{-\beta}\cos a + e^{-2\beta}}, \qquad A(n) = \frac{1}{\sqrt{1-2e^{-\beta}\cos\alpha+e^{-2\beta}}},$$

$\xi_n = \arccos \dfrac{1-e^{-\beta}\cos\alpha}{\sqrt{1-2e^{-\beta}\cos a + e^{-2\beta}}}$. In this specific case at $\alpha \ll 1$, $\beta \ll 1$ the amplitude

$$A(n) \simeq \frac{1}{\sqrt{\alpha^2 + \beta^2}}, \qquad \xi_n \simeq \arccos \frac{\beta}{\sqrt{\alpha^2 + \beta^2}}. \tag{A3}$$

A transient behavior of the amplitude $A_n$ and the relative power ($P_n = A_n^2$) for the time interval $\tau \simeq T_1/\beta$ ($n_\tau = 1/\beta$) is presented on the Fig. 4. At this interval the amplitude and the emitted power can be much higher then theirs average meanings. That is why it will be useful to extract periodically the energy of the light beam in the optical resonator with the period $T_{extr} \simeq \tau$ or to change the phase of the stored radiation in the URWs to $\pi$ for one revolution of the beam in the ring (production of the overload conditions).

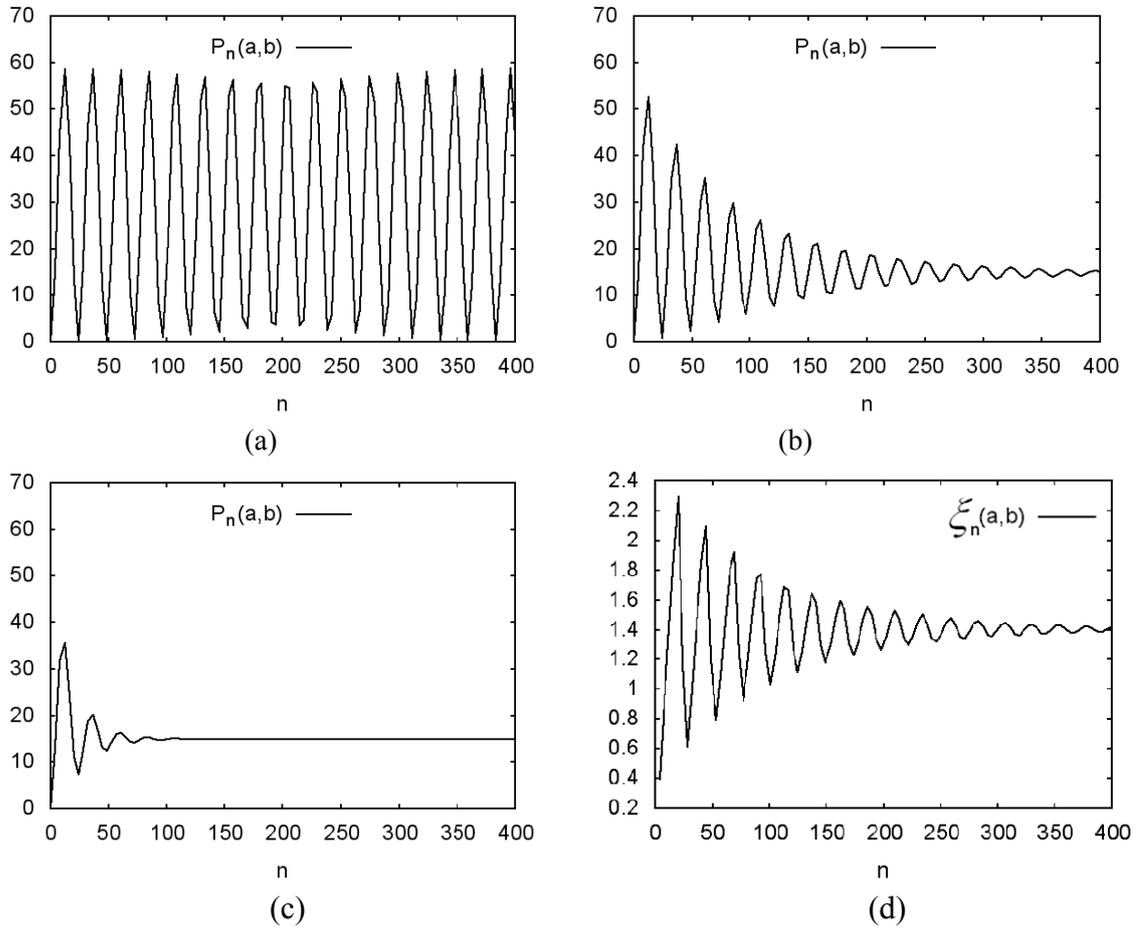

Figure 4: Time dependence of the SSUR power emitted by a particle at $\alpha = \pi/12$, $\beta = 0$ (a); $\alpha = \pi/12$, $\beta = 0.01$ (b); $\alpha = \pi/12$, $\beta = 0.05$ (c), $\alpha = \pi/12$, $\beta = 0.01$ (d), $n_c \gg n_\tau$.